\documentstyle[preprint,aps]{revtex}
\begin{document}
\draft

\title{Nucleon-nucleon elastic scattering analysis to 2.5 GeV}
\author{Richard A. Arndt, Chang Heon Oh, 
Igor I. Strakovsky$^\dagger$ and Ron L. Workman}
\address{Department of Physics, Virginia Polytechnic Institute and State
University, Blacksburg, VA~24061}
\author{Frank Dohrmann} 
\address{I. Institut f\"ur Experimentalphysik, Universit\"at Hamburg,
Luruper Chaussee 149, D-22761 Hamburg}

\date{\today}
\maketitle

\begin{abstract}

A partial-wave analysis of $NN$ elastic scattering data has been completed. 
This analysis covers an expanded energy range, from threshold to a 
laboratory kinetic energy of 2.5 GeV, in order to include recent elastic 
$pp$ scattering data from the EDDA collaboration. The results of both 
single-energy and energy-dependent analyses are described. 
				  
\end{abstract}


\centerline{I. INTRODUCTION}
\vskip .2cm

This analysis of elastic nucleon-nucleon scattering data updates our
previous analysis\cite{vpi94} to 1.6 GeV in the laboratory kinetic energy. 
The present analysis extends to 2.5 GeV, which is the limit for elastic
$pp$ differential cross sections measured\cite{edda} 
by the EDDA collaboration using the cooler synchrotron at COSY. 

Measurements with a laboratory kinetic energy near 2 GeV are particularly 
interesting as they correspond to a center-of-mass energy (2.7 GeV) which
has been suggested\cite{dibaryon} for a dibaryon resonance\cite{dp}. 
Near this energy, a sharp 
structure has been reported in the polarization observable 
$A_{yy}$\cite{ball}, and this was taken as support for such a resonance. 
A resonancelike structure, at about the same energy, has also been reported
in an analysis by Hoshizaki\cite{hosh}.
The authors of Ref.\cite{edda} have considered this possibility, 
but find no evidence for a resonant excursion in their cross sections. 
Polarization measurements expected from COSY and SATURNE II 
will certainly help to clarify this issue. 

The data base above 1.6 GeV is mainly comprised of cross section 
measurements, much of this coming from Ref.\cite{edda}. In Section~II 
we describe the expanded database, noting the additions 
below 1.6 GeV as well as the new region from 1.6 GeV to 2.5 GeV. 
While the most significant changes are seen in our $pp$ partial wave
amplitudes, both $pp$ and $np$ data have been analyzed.

In Section~III, we briefly review the formalism used in our analyses. 
Here we present the updated amplitudes and make comparisons with our
previous solution (SM94)\cite{vpi94}. 
Fits with and without the new EDDA data are  
compared to show the influence of this particular measurement. Representative 
plots showing the agreement between our analysis (SM97) and cross section 
data have been generated to illustrate the quality of this fit.
These results and the prospect for improvements are summarized in Section~IV. 

\vskip .5cm
\centerline{II. THE DATABASE}       
\vskip .2cm

Our previous $NN$ scattering analyses\cite{vpi94} were based on 12838 $pp$ 
and 10918 $np$ data.  In Ref.\cite{vpi94} the $pp$ analysis extended up to a 
laboratory kinetic energy of 1.6~GeV; the $np$ analysis was truncated at 
1.3~GeV.  The present database\cite{said} 
is considerably larger due both to an expanded 
energy range for the $pp$ system and the addition of new data at lower
energies. The total database is now about 20\% larger than was used
in our previous analysis\cite{vpi94}. 

Below we list recent additions to
our database. Some data sets which we collect are not used in the analyses, 
but are retained so that comparisons can be made. A complete description of 
the database and those data not included in our analyses is available from
the authors\cite{said}.  

The new $pp$ data have been produced mainly at COSY\cite{edda}.
From this source, we have added differential cross sections ranging from 
540 MeV to 2520~MeV in the proton kinetic energy and from 35$^\circ$ to 
90$^{\circ}$ in the cm scattering angle. In addition to this, about 60 high 
quality polarized data (P, A$_{xx}$, A$_{yy}$, and A$_{zx}$) at 200~MeV were 
produced by the Indiana cooler\cite{HA97}.  Another 35 high accuracy 
differential cross sections between 490 and 790~MeV were recently 
published\cite{SI96}. These measurements were made at LAMPF.  
We have added an excitation function of cross sections 
at 90$^{\circ}$ and between 0.3 and 0.4~MeV. These were measured at the
M\"unster University low-energy machine\cite{DO97P}.  
We have also added a measurement of A$_{zz}$ at 650~MeV produced by 
LAMPF\cite{PA88} but missed in the SAID database\cite{vpi94}.  

In constructing the data base extension from 1600 MeV to 2500 MeV, we
reexamined a number of references in order to include higher energy data
which had previously been neglected.  
This search netted additional data mainly from ANL (450 points) and 
Saclay (893 points).  The complete set is listed\cite{AB751} --
\cite{WI72} in alphabetical order.

The $np$ database has not been increased significantly and, as 
a result, we did not extend our analysis of the I~=~0 system. 
New $np$ polarized data have been produced mainly by TRIUMF 
(101 points)\cite{DA96,BA95}, 
IUCF (33 points)\cite{BO94P}, and LAMPF (49 points)\cite{MC96}.  
The  ANL--LAMPF--New Mexico 
University--Texas A \& M University collaboration has finalized its analysis 
of 311 high quality $np$ polarized observables (A$_{xx}$, A$_{zz}$, A$_{yy}$, 
and A$_{zx}$) between 485 and 790~MeV and ranging from 25$^\circ$ 
to 180$^{\circ}$ \cite{CA96}.  
These measurements were published previously in Ref.\cite{GA89}.  
A few total cross sections in pure spin states between 4 and 16~MeV were 
produced by TUNL\cite{WI95} and Charles University 
at Prague\cite{BR96,DO96P}.  Recently, the final LAMPF $\Delta \sigma _{L}$ 
measurements between 480 and 790~MeV were also published\cite{BE94}.  
In addition, some new $\Delta \sigma _{L}$ measurements 
above 1190~MeV were made at JINR (Dubna)\cite{AD96}.
Added unpolarized measurements include 15 $np$ differential cross sections 
at 67~MeV from PSI\cite{GO94}
and 6 differential cross sections at 14~MeV from T\"ubingen University
\cite{ME97}. A few missed differential 
cross sections at low energies from LAMPF\cite{DR78P} and at 1240~MeV
from Berkeley\cite{PE70} were also added.

A few data sets were added to the data base but not included in the analysis.
These include 82 missed $np$ total cross section
measurements between 4 and 231~MeV from LAMPF\cite{LI80P}.  We excluded 
these data from the analysis in order to retain the same database 
(below 350~MeV) as was used in the Nijmegen analysis\cite{ni94}.  
This also applies to a new set of $np$ differential cross
sections at 162~MeV and at backward angles which were measured at the 
Svedberg Facility at Uppsala\cite{ER95}.

\vskip .5cm
\centerline{III. PARTIAL-WAVE ANALYSIS}
\vskip .2cm

Our first attempts to extend the range of the $NN$ analysis
used the parameterization scheme of Ref.\cite{vpi94}. These were 
unsuccessful. The problem was traced to the basis functions used to
expand our K-matrix elements. Many of these become nearly degenerate as the
kinetic energy of the incoming nucleon ($T$) increases to 2.5 GeV. 
As a result, a modified form
was used in the present analysis. Apart from this difference, the 
formalism used here is identical to that used in Ref.\cite{vpi94}. The
reader is directed to Refs.\cite{ar87,maw} for more details. In the 
following we just outline the method used, in order to show how the
modified basis functions fit into our parameterization scheme. 

For uncoupled partial waves 
($^1D_2$, $^3F_3$, ...), an S-matrix ($S=S_E S_I$) is used.
This product S-matrix is constructed from exchange ($S_E$) and inelastic 
($S_I$) pieces. $S_E$ is parameterized in terms of a K-matrix
\begin{equation}
S_E \; = \; \left( 1+iK_E \right) /  \left( 1-iK_E \right)  ,
\end{equation}
which in turn is expanded as
\begin{equation}
K_E \; = \; {\rm Born} \; + \: \sum_i \alpha_i A_{li}  .
\end{equation}
Here the Born term gives the single-pion exchange contribution and 
$\alpha_i$ are free parameters. 
The expansion basis elements, $A_{li}$, are given by
\begin{equation}
A_{li} \; = \; F_{li} \left( { {T} \over { T + T_C} } \right)^{i-1}  ,
\end{equation}
where the function $F_{li}$, used as the expansion basis in our previous
fits\cite{vpi94}, is given by
\begin{equation}
F_{li} (T) \; = \; {{4\mu^2}\over {MT} } \int^1_0 Q_l \left( {{x_0 - x}\over
 {1-x}} \right) {{x^{i-1/2} }\over {1-x} } dx .
\end{equation}
Here $M$ ($\mu$) is the nucleon (pion) mass and 
$x_0 = 1 + (4\mu^2/MT)$. $Q_l$ is a Legendre function of the second kind. 
In Eq.~(3), $T_C$ is a parameter which was chosen to be 1 GeV. (The 
fit was not sensitive to this choice; fits using 0.5 GeV and 1.5 GeV were
also attempted.)
The basis function given in Eq.~(4) was derived in Ref.\cite{maw}.

To ensure time-reversal invariance, the spin-coupled waves (for example,
$^3P_2$ $-$ $^3F_2$) are parameterized as
\begin{equation}
S(2\times 2) \; = \; S^{1/2}_E S_I S^{1/2}_E  ,
\end{equation}
where again the matrix $S_E$ is expanded in terms of a K-matrix
with the elements
\begin{equation}
K_m \; = \; {\rm Born}_m + \sum_i A_{l_m i} ,
\end{equation}
the subscript ($m=(+,0,-)$) labeling states with $l_m = (J+1, J, J-1)$. 
As in Ref.\cite{vpi94}, the matrix $S_I$ is taken from a Chew-Mandelstam
K-matrix coupling the $NN$ channel to an appropriate $N\Delta$ state.
This has been extensively described in Ref.\cite{ar87}.
The simple modification of the basis elements, displayed in Eq.~(3), 
provided the added flexibility required to extend our analysis to 2.5 GeV. 

In Table~I, we compare the energy-dependent and single-energy fits over
the energy bins used in the single-energy analyses. Also listed are the 
number of parameters varied in each single-energy solution. A total of
144 parameters were varied in the energy-dependent analysis. 

Our single-energy and energy-dependent results for  the isovector and
isoscalar partial-wave amplitudes are displayed in Figs.~1 and 2. 
Here we also compare with our previous fit (SM94). In some cases the
changes are quite large. This is particularly true near the upper energy
limit of SM94, and for the smaller partial waves. The effect of these
changes can be clearly seen in Fig.~3, where we show how well the new
EDDA data\cite{edda} are reproduced by both SM94 and SM97. The influence 
of this experiment is most pronounced in the forward direction. 

In general,
we find little structure over the higher energy region. This reflects the
smooth, and rather flat, total and reaction cross sections between 1.5 GeV
and 2.5 GeV. Our fit to these quantities is displayed in Fig.~4. Note that
the reaction cross sections were excluded from our fit. This verifies that
the set of total, total elastic (deduced from differential cross sections), 
and reaction cross sections are self-consistent. 

The present analysis actually gives an improved fit to the data below 
1.6 GeV. This is due to the altered basis set, found necessary to fit
the higher energy data. Numerical comparisons are given in Table~II. 
Here we see that the COSY data\cite{edda} comprise a large fraction 
of the total set above 1.6 GeV. The results of analyses with (SM97) 
and without (NM97) this data set show how influential these measurements 
have been in determining the amplitudes. (The fits SM97 and NM97 used
identical parameterization schemes. Only the data base was changed.) 
The COSY data contribute a
$\chi^2$/datum of 1.07 when included in the fit. This jumps to 5.6 
when we attempt a prediction based on the remaining data.

\vskip .5cm
\centerline{IV. CONCLUSIONS AND FUTURE PROSPECTS}
\vskip .2cm

We have extended our $pp$ partial-wave analyses nearly 1 GeV beyond
the limit quoted in our previously published results\cite{vpi94}.
The present range has been selected to include all of the recent
elastic $pp$ cross section data measured by the EDDA group\cite{edda}.
We found that it was possible to simultaneously fit the $pp$ total
cross section data, in particular the precise data of Ref.\cite{BU66}, 
along with differential cross sections from the
EDDA collaboration\cite{edda}. The resulting reaction cross sections,
which were not included in the fit, are quite well reproduced.
The predicted reaction cross sections are consistent with the results of
Ref.\cite{verw} at lower energies, but deviate from these and follow the
results of Ref.\cite{sgtr} above 1 GeV. 

While we find that the partial-wave amplitudes above 1.6 GeV are
smooth and structureless, reflecting the behavior seen in the total
and elastic cross section data, we have also considered the effect of more
localized structures reported in polarization 
measurements\cite{dibaryon,ball}.
We can add resonancelike structures in individual partial-waves to see
their effect on any observable. This will be utilized as more polarization
data become available. 

As the high energy region was constrained mainly by cross section
data, the present solution should be considered as a guide to the
expected amplitudes. 
The EDDA collaboration is planning to measure 
P, A$_{yy}$, A$_{xx}$, and A$_{xz}$ in the near future.
This will be crucial to any future analyses. 

Further data is also expected from a number of other labs. 
About 2000 polarized $pp$ measurements are expected
above 1000~MeV\cite{i1} as the nucleon-nucleon program at SATURNE II
is completed. While not included in the present fit, preliminary 
data\cite{ar97} from SATURNE II is in reasonable agreement with our 
predictions. A representative fit to P data, 
at 2.16 GeV, is given in Fig.~5.

A similar number of polarized quantities from $np$ elastic scattering 
are expected (between 250 and 560~MeV\cite{i3}) from PSI.
The Freiburg University group is also planning to publish $np$ measurements 
which were done at PSI at the beginning of the 1980's.  These data range 
from 200 to 580~MeV and from 77$^{\circ}$ to 179$^{\circ}$\cite{i4}.
Final $np$ differential cross sections between 73$^{\circ}$ 
and 179$^{\circ}$ measured at Uppsalla\cite{i5} are expected to replace 
data at 96~MeV\cite{RO92} and 162~MeV\cite{ER95}. 
IUCF is also measuring $np$ differential cross sections in the backward 
direction at about 200~MeV to solve a shape 
problem in the angular distribution\cite{i6}.
Other $np$ sources include an extension of 
$\Delta \sigma _{L}$ measurements\cite{i7} at JINR\cite{AD96},
TRIUMF analyzing power measurements at 
350~MeV\cite{i8}, 
and TUNL measurements\cite{i9} of the P parameter and ${\Delta \sigma}_L$ 
at 7 and 15~MeV.  
We will continue to update our energy-dependent and single-energy solutions
as the new measurements become available. 

Finally we note that by extending our analysis to 2.5 GeV, we may be bridging
the gap between the low- and high-energy regions. This is suggested if we 
plot $d\sigma /dt$ versus $s$, as is shown in Fig.~6. The result expected
from dimensional counting at high-energy and fixed 
cm angle\cite{brodsky,farrar} is
\begin{equation}
{ {d\sigma} \over {d t} } \sim { {1}\over {s^{N-2}} } = s^{-10}   ,
\end{equation}
where $N$ is the minimum number of fundamental constituents (quarks). 
While a slightly extended energy range would be more definitive, our results
do appear to be consistent with this limit. 

\vskip .5cm
\centerline{ACKNOWLEDGMENTS} 
\vskip .2cm

The authors express their gratitude to 
M. Bachman, J. Ball, D. V. Bugg, W. B\"urkle, C. A. Davis,  
Z. Dolezal, H. Dombrowski, M. Drosg, T. E. O. Ericson, 
G. Glass, W. Haeberli, J. Jourdan, 
F. Lehar, C. Lechanoine-LeLuc, P. W. Lisowski, 
B. Loiseau, M. J. McNaughton, D. F. Measday, 
G. Mertens, N. Olsson, H. Rohdje\ss{}, V. I. Sharov, H. M. Spinka, 
W. Tornow, S. Vigdor, and W. S. Wilburn  
for providing experimental data prior to publication or for clarification of 
information already published. We also thank B. Z. Kopeliovich for helpful
discussions. I.~S. acknowledges the hospitality extended by the 
Physics Department of Virginia Tech. F.~D. would like to thank the 
EDDA collaboration for their support. 
This work was supported in part by the U.~S.~Department of Energy Grant
DE--FG05--88ER40454. F.~D. was supported by BMBF contract 06HH561Tp2. 

\newpage


\eject


{\Large\bf FIGURE CAPTIONS}\\
\newcounter{fig}
\begin{list} {Figure \arabic{fig}.}
{\usecounter{fig}\setlength{\rightmargin}{\leftmargin}}
\item
{Isovector partial-wave amplitudes from 0 to 2.4 GeV in the proton 
kinetic energy. Solid curves give the 
amplitudes corresponding to the SM97 solution. The real (imaginary) parts of
the single-energy solutions are plotted as triangles (squares). For 
comparison, the previous solution SM94\cite{vpi94} is plotted with (+) marks. 
The (x) marks give \hbox{Im$T$-$T^2$-$T_{sf}^2$} from SM97, 
where $T_{sf}^2$ is the spin-flip amplitude. 
All amplitudes are dimensionless.}
\item
{Isoscalar partial-wave amplitudes from 0 to 1.2 GeV. Notation as in Fig.~1.}
\item
{Comparison between SM97 (solid curve) and differential cross sections at
 (a) $\theta^* = 45^\circ \pm 1^\circ$ and 
 (b) $\theta^* = 90^\circ \pm 1^\circ$. Recent COSY measurements\cite{edda} 
 are plotted as filled circles. Other data from the SAID data 
 base\cite{said} are plotted as
 crosses. Our previous solution (SM94) is plotted to 1.6 GeV (dot-dashed
 line). }
\item
{Total cross section comparisons. (a) The solid (dashed) curves give  
the predictions of solution SM97 for the total (total elastic) cross section.
Experimental points are from the SAID data base\cite{said}; 
filled circles are from Ref.\cite{BU66}. (b) The solid curve gives the
total reaction cross section of SM97. Filled circles are estimates from
Ref.\cite{verw}. Filled triangles are estimates from Ref.\cite{sgtr}.}
\item
{Angular dependence of recent SATURNE II analyzing power ($P$)
data\cite{ar97}. This measurement, at 2.16 GeV, was not included the SM97
analysis. The solid line gives the SM97 prediction. The dashed lines are
generated from a single-energy solution and its associated error estimate.}
\item
{$d\sigma /dt$ plotted as a function of $s$ at $\theta^* = 90^\circ$. The
SM97 solution is plotted as a solid curve. The dash-dotted line gives
$d\sigma /dt \sim s^{-10}$. The plotted data are from Ref.\cite{ak67}.}
\end{list}

\eject


Table~I. Comparison of the single-energy (SES) 
and energy-dependent (SM97) fits
to $pp$ and $np$ data. Values of $\chi^2$ are given for the 
SES and SM97 fits (evaluated over the same energy bins). Also listed
is the number of parameters varied in each single-energy solution.
 
\vskip 10pt
\centerline{
\vbox{\offinterlineskip
\hrule
\hrule
\halign{\hfill#\hfill&\qquad\hfill#\hfill&\qquad\hfill#\hfill
 &\qquad\hfill#\hfill \cr
\noalign{\vskip 6pt}
Energy Range (MeV) & $\chi^2$ SES(SM97)/$pp$ data 
 & $\chi^2$ SES(SM97)/$np$ data & Parameters  \cr
\noalign{\vskip 6pt}
\noalign{\hrule}
\noalign{\vskip 10pt}
 4-6     &   22(39)/28   & 50(66)/53 & 6 \cr
\noalign{\vskip 6pt}
 7-12    &   84(132)/88  & 221(309)/87 & 6 \cr
\noalign{\vskip 6pt}    
 11-19   &   17(49)/27   &   191(445)/236  &  8 \cr
\noalign{\vskip 6pt}    
 19-30   &   123(275)/114   &   263(286)/295  &   8  \cr
\noalign{\vskip 6pt}    
 32-67   &   294(375)/224   &   667(754)/485  &   10   \cr
\noalign{\vskip 6pt}    
 60-90   &   55(64)/72      &   457(595)/329  &  10   \cr
\noalign{\vskip 6pt}   
 80-120  &   161(185)/154   &   419(487)/353  &  10   \cr
\noalign{\vskip 6pt}   
 125-174 &   301(310)/287   &   328(367)/272  &   11   \cr
\noalign{\vskip 6pt}   
 175-225  &  249(354)/212   &   715(766)/499  &   13   \cr
\noalign{\vskip 6pt} 
 225-270  &  66(91)/64      &   243(270)/236  &   13   \cr
\noalign{\vskip 6pt}   
 276-325  &  274(309)/256   &   571(655)/518  &   17   \cr
\noalign{\vskip 6pt}   
 325-375 &   297(320)/246   &   421(474)/353  &   17   \cr
\noalign{\vskip 6pt}   
 375-425 &   555(601)/436   &   753(843)/549  &   17   \cr
\noalign{\vskip 6pt}   
 425-475 &   902(1004)/665  &   775(799)/629  &   18   \cr
\noalign{\vskip 6pt}   
 475-525 &   1322(1484)/1081  &   1252(1419)/787  &   30    \cr
\noalign{\vskip 6pt}   
 525-575 &   861(972)/754    &   549(584)/432  &   31    \cr
\noalign{\vskip 6pt}   
 575-625 &   1032(1154)/760  &   422(491)/367   &   36   \cr
\noalign{\vskip 6pt}   
 625-675 &   891(863)/754    &   1263(1563)/875  &   36   \cr
\noalign{\vskip 6pt}   
 675-725 &   838(882)/777   &    403(473)/386    &   37    \cr
\noalign{\vskip 10pt}}
\hrule
\hrule}}  
\vfill

\eject

Table~I. (continued) 

\vskip 10pt
\centerline{
\vbox{\offinterlineskip
\hrule
\hrule
\halign{\hfill#\hfill&\qquad\hfill#\hfill&\qquad\hfill#\hfill
 &\qquad\hfill#\hfill \cr
\noalign{\vskip 6pt}  
Energy Range (MeV) & $\chi^2$ SES(SM97)/$pp$ data 
 & $\chi^2$ SES(SM97)/$np$ data & Parameters  \cr
\noalign{\vskip 6pt}
\noalign{\hrule}
\noalign{\vskip 10pt}
 725-775 &   990(1195)/827  &    512(558)/374   &   37    \cr
\noalign{\vskip 6pt}   
 775-824 &   1583(1754)/1170  &  1518(1845)/944  &   38    \cr
\noalign{\vskip 6pt}   
 827-874  &  1195(1358)/939   &  386(497)/366  &   39    \cr
\noalign{\vskip 6pt}   
 876-924  &  341(412)/389     &  753(920)/628   &   41   \cr
\noalign{\vskip 6pt}   
 926-974 &   790(945)/679     &  354(498)/352   &   43   \cr
\noalign{\vskip 6pt}   
 976-1020 &  931(1131)/708  &    300(441)/331   &   43    \cr
\noalign{\vskip 6pt}
 1078-1125 &   528( 689)/413  &  427(671)/326   &    45   \cr
\noalign{\vskip 6pt}  
 1261-1299 &   680(972)/507   &      $---$      &    29   \cr
\noalign{\vskip 6pt} 
 1481-1521 &   139(266)/149   &      $---$      &    29    \cr
\noalign{\vskip 6pt}  
 1590-1656 &   472(655)/409  &       $---$      &    31    \cr
\noalign{\vskip 6pt}  
 1685-1724  &  185(293)/118  &       $---$      &    31   \cr
\noalign{\vskip 6pt}  
 1778-1818 &   404(628)/347  &       $---$      &    31   \cr
\noalign{\vskip 6pt}  
 1929-1968  &  218(271)/168  &       $---$      &    31   \cr
\noalign{\vskip 6pt}  
 2065-2104 &   673(1241)/431   &     $---$      &    31   \cr
\noalign{\vskip 6pt}  
 2176-2224 &   1005(1325)/377  &     $---$      &    31   \cr
\noalign{\vskip 6pt}  
 2330-2470 &   803(1257)/458  &      $---$      &    31   \cr
\noalign{\vskip 10pt}}
\hrule
\hrule}}  
\vfill
\eject
Table~II. Comparison of present and previous solutions. Dataset A was used 
in the SM94 analysis\cite{vpi94}. Dataset B contains all data (apart from
the EDDA data\cite{edda}) used in generating solution SM97. See the text for
details regarding the SM97 and NM97 fits.

\vskip 10pt
\centerline{
\vbox{\offinterlineskip
\hrule
\hrule
\halign{\hfill#\hfill&\qquad\hfill#\hfill&\qquad\hfill#\hfill 
 &\qquad\hfill#\hfill \cr
\noalign{\vskip 6pt}
PWA & Data & $\chi^2$/$pp$ data & $\chi^2$/$np$ data \cr
\noalign{\vskip 6pt}
\noalign{\hrule}
\noalign{\vskip 6pt}  
  & & (0-1600 MeV)  &  (0-1300 MeV)  \cr
\noalign{\vskip 6pt}
\noalign{\hrule}
\noalign{\vskip 10pt}
SM94\cite{vpi94} & (dataset A)   &   22375/12838  & 17516/10918 \cr
\noalign{\vskip 6pt}
SM94\cite{vpi94} & (dataset B)   &   22390/12889  & 18480/10843 \cr
\noalign{\vskip 6pt}  
SM97 & (dataset B)   &   20910/12889  & 17400/10843 \cr
\noalign{\vskip 6pt}  
\noalign{\hrule}
\noalign{\vskip 6pt}   
   & &  (0-2520 MeV) &  (0-2000 MeV)  \cr
\noalign{\vskip 6pt}
\noalign{\hrule}
\noalign{\vskip 10pt}  
SM97 & (dataset B)   &  26460/14873  &  17440/10854 \cr
\noalign{\vskip 6pt}  
SM97 & (EDDA dataset\cite{edda}) & 2278/2121 & $-$ \cr
\noalign{\vskip 6pt}  
NM97 & (dataset B)  &  25240/14873  & 17280/10854 \cr
\noalign{\vskip 6pt}  
NM97 & (EDDA dataset\cite{edda}) & 11964/2121 & $-$ \cr
\noalign{\vskip 10pt}}
\hrule
\hrule}}  
\vfill
\eject

\end{document}